\documentclass[12pt]{article}%
\usepackage{amsmath}
\usepackage{amsfonts}
\usepackage{amssymb}
\usepackage{graphicx}%
\providecommand{\U}[1]{\protect\rule{.1in}{.1in}}
\begin{document}
\begin{titlepage}
\ \\
\begin{center}
\LARGE
{\bf
Quantum Energy Teleportation \\
with Trapped Ions
}
\end{center}
\ \\
\begin{center}
\large{
Masahiro Hotta
}\\
\ \\
\ \\
{\it
Department of Physics, Faculty of Science, Tohoku University,\\
Sendai, 980-8578, Japan\\
hotta@tuhep.phys.tohoku.ac.jp
}
\end{center}
\begin{abstract}
We analyse a protocol of quantum energy teleportation that
transports energy from the left edge of a linear ion crystal to the
right edge by local operations and classical communication at a speed
considerably greater than the speed of a phonon in the crystal. A
probe qubit is strongly coupled with phonon fluctuation in the ground
state for a short time, and it is projectively measured in order to
obtain information about this phonon fluctuation. During the measurement
process, phonons are excited by the time-dependent measurement interaction,
and the energy of the excited phonons must be infused from outside
the system. The obtained information is transferred to the right
edge of the crystal through a classical channel. Even though the phonons
excited at the left edge do not arrive at the right edge at the same
time as when the information arrives at the right edge, we are able
to soon extract energy from the ions at the right edge by using the
transferred information. Because the intermediate ions of the crystal
are not excited during the execution of the protocol, energy is transmitted
in the energy transfer channel without heat generation.
\end{abstract}
\end{titlepage}

\bigskip

\bigskip

\section{Introduction}

\bigskip{}

~

A trapped ion system is expected to be a promising candidate for developing
quantum computers \cite{CZ}. Recently, the experimental development of trapped
ions has undergone a significant technical advancement \cite{HRB}.\ In
addition, this technology has been applied to quantum teleportation
\cite{tele}\ and quantum error correction \cite{qec}. The successful
application of trapped ion systems to the abovementioned tasks proves that
these systems have a great potential for use in other fascinating quantum
tasks, which have not yet been experimentally executed. In this study, we
analyse quantum energy teleportation with cold ions in a linear trap, that is,
a linear ion crystal. Recently, quantum energy teleportation (QET) has been
proposed \cite{hotta1}-\cite{hotta4}; this protocol is able to transport
energy simply by local operations and classical communication. Energy can be
effectively teleported without breaking any physical laws including causality
and local energy conservation by measuring zero-point oscillation at one site
in the entangled ground state of a many-body quantum system and providing the
measurement result to distant sites. The key point is that there exists
quantum correlation between local fluctuations of different sites in the
ground state. Therefore, the measurement result of local fluctuation in some
site includes information about fluctuation in other sites. By selecting and
performing a suitable local operation based on the transferred information,
the zero-point oscillation of a site away from the measurement site can be
more suppressed than that of the ground state, yielding negative energy
density. Here, the origin of energy density is fixed such that the expectation
value vanishes for the ground state. It is well known that such regions with
negative energy density are created by superposing energy eigenstates quantum
mechanically. Even if a system comprises a region with negative energy
density, other regions with positive energy density still exist, and
therefore, the total energy of the system remains non-negative. Discussions on
negative energy density of non-relativistic systems are given in \cite{hotta3}
and those for negative energy density of relativistic systems are given in
\cite{BD}. While performing the above local operation to create negative
energy density in the system, the surplus energy is transferred from quantum
fluctuations to external systems and can be harnessed. \ Though the protocols
of energy teleportation can be implemented for other systems such as quantum
fields \cite{hotta1} and spin chains \cite{hotta2,hotta3,hotta4}, linear ion
crystals can be effectively used for QET. Due to the Coulomb interaction and
the harmonic potential for linear trapping, the ions interact with each other
so strongly that entanglement for energy teleportation is easily generated in
the phonon ground state. It is also possible to POVM measure the local
fluctuation of phonons in the linear ion crystal by coupling the phonon modes
with the internal energy levels of each ion via a laser field \cite{HRB}.

In this study, we use the linear trap models that are extensively analysed by
James \cite{James} and consider a protocol of QET. In what follows, $N$ cold
ions, which are strongly bound in the $y$ and $z$ directions but weakly bound
in a harmonic potential in the $x$ direction, form a linear ion crystal called
the QET channel. The first ion that stays at the left edge of the crystal is
the gateway of the QET channel where energy is input. The $N$-th ion that
stays at the right edge of the crystal is the exit of the QET channel where
the teleported energy is output. We select two suitable internal energy levels
of the gateway ion and regard them as the energy levels of a probe qubit in
order to measure the local phonon fluctuation. The probe qubit is strongly
coupled with the phonon fluctuation in the ground state for a short period of
time via a laser field, and it is projectively measured in order to obtain
1-bit information about phonon fluctuation. During the measurement process at
the gateway, phonons are excited in the system by the time-dependent
measurement interaction, and hence, the energy of the excited phonons must be
infused from outside the system. In the measurement models used in this study,
the kinetic energy of the gateway ion increases after the measurement;
however, the kinetic energy of other ions and the potential energy of all the
ions remain unchanged. This infusion of energy is regarded as energy input at
the gateway of the QET channel. The obtained information is transferred
through a classical channel from the gateway point to the exit point. In
principle, the speed at which this information is transferred is equal to the
speed of light, which is considerably greater than that of phonon
propagation\ in the ion crystal. It is emphasized that even when the phonons
excited at the QET gateway do not arrive at the exit point, the information
still arrives at the exit point. Because the measurement process at the
gateway is local, the state of the exit ion before the phonons arrive is
locally the same as the ground state. Surprisingly, however, we are able to
soon extract energy from the exit ion by using the transferred information.
Performing a local operation that depends on the measurement result suppresses
the zero-point oscillation of the phonon fluctuation at the exit ion, yielding
negative energy density around the exit. The surplus energy of the fluctuation
is transferred from the phonon system to the external systems including the
device, in order to perform this operation. Here, it should also be emphasized
that without the measurement result, we cannot extract energy from the exit
ion in the local ground state by performing an arbitrary local quantum
operation on the ion. Because the intermediate ions of the crystal are not
excited during the protocol execution, this QET ensures energy transportation
in the channel without the generation of heat. The non-negativity of the total
Hamiltonian of phonons ensures that the amount of the teleported energy is
less than the amount of the input energy. The amount of energy required for
transferring the information is negligibly small in principle as compared to
the amount of the teleported energy, because the information can be simply
transported by low-energy physical carriers such as electromagnetic waves with
long wavelengths.

The paper is organized as follows. In section\ 2, the linear ion crystal model
developed by James \cite{James} is reviewed. In section 3, QET for the model
is analysed. In the last section, the summary and discussion are given. In
this study, we adopt a unit $\hbar=1$.

\section{\ \ Linear Ion Crystal Model}

\ \newline~Following James \cite{James}, let us consider $N~$ions with charge
$Ze$ and mass $m$ that are strongly bound in the $y$ and $z$ directions but
weakly bound in a harmonic potential in the $x$ direction. The position of the
$n$-th ion is denoted by $x_{n}(t)$, where the ions are numbered from left to
right. The Hamiltonian is expressed as%
\begin{equation}
H_{ion}=\sum_{n=1}^{N}\frac{m}{2}\dot{x}_{n}(t)^{2}+\sum_{n=1}^{N}\frac{m}%
{2}\nu^{2}x_{n}(t)^{2}+\frac{1}{2}\sum_{\substack{n,n^{\prime}=1,\\n\neq
n^{\prime}}}^{N}\frac{Z^{2}e^{2}}{\left\vert x_{n}(t)-x_{n^{\prime}%
}(t)\right\vert }, \label{1}%
\end{equation}
where the dot denotes time derivative, and $\nu$ is the trap frequency, which
characterizes the strength of the trapping harmonic potential in the $x$
direction. In the case of cold ions, we can approximate the position of the
$n$-th ion by
\[
x_{n}(t)\approx x_{n}^{(0)}+q_{n}(t),
\]
where $x_{n}^{(0)}$ is the position of the ion in the ground state, and
$q_{n}(t)$ is a small displacement describing the phonon modes in the ion
crystal. The equilibrium positions in the ground state are determined by the
following equation derived from Eq. (\ref{1}):
\begin{equation}
m\nu^{2}x_{n}^{(0)}-\sum_{n^{\prime}=1}^{n-1}\frac{Z^{2}e^{2}}{\left(
x_{n}^{(0)}-x_{n^{\prime}}^{(0)}\right)  ^{2}}+\sum_{n^{\prime}=n+1}^{N}%
\frac{Z^{2}e^{2}}{\left(  x_{n}^{(0)}-x_{n^{\prime}}^{(0)}\right)  ^{2}}=0.
\label{2}%
\end{equation}
Let us introduce a scale length given by%
\[
l=\left(  \frac{Z^{2}e^{2}}{m\nu^{2}}\right)  ^{1/3}%
\]
and rescale the position variables as%
\[
u_{n}=\frac{x_{n}^{(0)}}{l}.
\]
Then, Eq. (\ref{2}) is rewritten in a dimensionless form as
\begin{equation}
u_{n}-\sum_{n^{\prime}=1}^{n-1}\frac{1}{\left(  u_{n}-u_{n^{\prime}}\right)
^{2}}+\sum_{n^{\prime}=n+1}^{N}\frac{1}{\left(  u_{n}-u_{n^{\prime}}\right)
^{2}}=0. \label{3}%
\end{equation}
First, it is pointed out \cite{James} that Eq. (\ref{3}) can be analytically
solved for the cases with $N=2\,,3$. When $N=2\,$\ , the solution of the above
equation can be expressed as%
\[
u_{1}=-\left(  \frac{1}{2}\right)  ^{\frac{2}{3}},u_{2}=\left(  \frac{1}%
{2}\right)  ^{\frac{2}{3}}.
\]
When $N=3\,$\ , the solution can be expressed as%
\[
u_{1}=-\left(  \frac{5}{4}\right)  ^{\frac{1}{3}},u_{2}=0,u_{3}=\left(
\frac{5}{4}\right)  ^{\frac{1}{3}}.
\]
For $N$ greater than 3, $u_{n}$ were numerically solved by James. The
Hamiltonian of phonons in the crystal is derived by expanding Eq. (\ref{1}) in
terms of $q_{n}(t)$ and by considering only bilinear terms; it is expressed as
follows:
\begin{equation}
H=\sum_{n=1}^{N}\frac{1}{2m}p_{n}^{2}+\sum_{n,n^{\prime}=1}^{N}\frac{1}{2}%
m\nu^{2}A_{nn^{\prime}}q_{n}q_{n^{\prime}}-E_{g}, \label{5}%
\end{equation}
where the zero-point energy $E_{g}$ is subtracted from the original form in
order to make the lowest eigenvalue of $H$ zero, and a real symmetric matrix
$A_{nn^{\prime}}$ is defined by%
\begin{equation}
A_{nn}=1+2\sum_{\substack{n^{\prime\prime}=1\\n^{\prime\prime}\neq n}%
}^{N}\frac{1}{\left\vert u_{n}-u_{n^{\prime\prime}}\right\vert ^{3}}
\label{70}%
\end{equation}
for $n=n^{\prime}$ and%
\[
A_{nn^{\prime}}=-2\frac{1}{\left\vert u_{n}-u_{n^{\prime}}\right\vert ^{3}}%
\]
for $n\neq n^{\prime}$. The eigenvectors $b_{n}^{(k)}(k=1,2,\cdots,N)$ of
$A_{nn^{\prime}}$ determined by%
\[
\sum_{n^{\prime}=1}^{N}A_{nn^{\prime}}b_{n^{\prime}}^{(k)}=\mu_{k}b_{n}^{(k)}%
\]
are real. In addition, the eigenvalues $\mu_{k}$ of $A_{nn^{\prime}}$ are real
and non-negative \cite{James}. The eigenvectors are numbered in the order of
increasing eigenvalues and normalized as
\[
\sum_{n=1}^{N}b_{n}^{(k)}b_{n}^{(k^{\prime})}=\delta_{kk^{\prime}}.
\]
The first and second eigenvectors are analytically obtained. For $k=1$, the
eigenvector is given by%
\[
b_{n}^{(1)}=\frac{1}{\sqrt{N}},
\]
with its eigenvalue $\mu_{1}=1$. For $k=2\,$, the eigenvector is given by%
\[
b_{n}^{(2)}=\frac{u_{n}}{\sqrt{\sum_{n^{\prime}=1}^{N}u_{n^{\prime}}^{2}}},
\]
with its eigenvalue $\mu_{2}=3$. Higher eigenvalues and eigenvectors are
numerically solved and listed in table 2 of \cite{James}. \ Next, let us
introduce the normal modes of phonons as%
\[
Q_{k}=\sum_{n=1}^{N}b_{n}^{(k)}q_{n}.
\]
Then, $H$ is diagonalized for the modes as
\[
H=\sum_{k=1}^{N}\left(  \frac{1}{2m}P_{k}^{2}+\frac{1}{2}m\nu^{2}\mu_{k}%
Q_{k}^{2}\right)  -E_{g},
\]
where $P_{k}$ is the conjugate momentum of $Q_{k}$ and $P_{k}(t)=m\dot{Q}%
_{k}(t)$. Let us introduce phonon creation and annihilation operators as%
\begin{align*}
a_{k}^{\dag}  &  =\sqrt{\frac{1}{2m\nu\sqrt{\mu_{k}}}}P_{k}+i\sqrt{\frac
{m\nu\sqrt{\mu_{k}}}{2}}Q_{k},\\
a_{k}  &  =\sqrt{\frac{1}{2m\nu\sqrt{\mu_{k}}}}P_{k}-i\sqrt{\frac{m\nu
\sqrt{\mu_{k}}}{2}}Q_{k}.
\end{align*}
These operators satisfy the following commutation relation:
\[
\left[  a_{k},~a_{k^{\prime}}^{\dag}\right]  =\delta_{kk^{\prime}}.
\]
The ground state $|g\rangle$ of the phonon system is defined by
\begin{equation}
a_{k}|g\rangle=0, \label{100}%
\end{equation}
and simultaneously satisfies
\begin{equation}
H|g\rangle=0 \label{6}%
\end{equation}
due to the subtraction of the zero-point energy $E_{g}$. From Eq. (\ref{6}),
it is verified that $H$ is a non-negative operator. i.e. $H\geq0$. By using
the normal modes, it is easy to solve the Heisenberg operator for the
displacement of the $n$-th ion from the equilibrium position as%
\begin{align}
q_{n}(t)  &  =\sum_{k=1}^{N}b_{n}^{(k)}Q_{k}(t)\nonumber\\
&  =\sum_{k=1}^{N}b_{n}^{(k)}\frac{i}{\sqrt{2m\nu\sqrt{\mu_{k}}}}\left(
a_{k}e^{-i\nu\sqrt{\mu_{k}}t}-a_{k}^{\dag}e^{i\nu\sqrt{\mu_{k}}t}\right)  .
\label{8}%
\end{align}
Its corresponding conjugate momentum operator is solved as
\begin{align}
p_{n}(t)  &  =\sum_{k=1}^{N}b_{n}^{(k)}P_{k}(t)\nonumber\\
&  =\sum_{k=1}^{N}b_{n}^{(k)}\sqrt{\frac{m\nu\sqrt{\mu_{k}}}{2}}\left(
a_{k}e^{-i\nu\sqrt{\mu_{k}}t}+a_{k}^{\dag}e^{i\nu\sqrt{\mu_{k}}t}\right)  .
\label{9}%
\end{align}
These Heisenberg operators can be computed from their Schr\H{o}dinger
operators $q_{n}(=q_{n}(0))$ and $p_{n}(=p_{n}(0))$ as follows:%
\begin{align*}
q_{n}(t)  &  =\sum_{n^{\prime}=1}^{N}W_{nn^{\prime}}^{(1)}(t)q_{n^{\prime}%
}+\frac{1}{m\nu}\sum_{n^{\prime}=1}^{N}W_{nn^{\prime}}^{(2)}(t)p_{n^{\prime}%
},\\
p_{n}(t)  &  =-m\nu\sum_{n^{\prime}=1}^{N}W_{nn^{\prime}}^{(3)}(t)q_{n^{\prime
}}+\sum_{n^{\prime}=1}^{N}W_{nn^{\prime}}^{(1)}(t)p_{n^{\prime}},
\end{align*}
where $W_{nn^{\prime}}^{(r)}(t)~(r=1,2,3)$ are real symmetric matrices defined
by%
\begin{align*}
W_{nn^{\prime}}^{(1)}(t)  &  =\sum_{k=1}^{N}\cos\left(  \nu\sqrt{\mu_{k}%
}t\right)  b_{n}^{(k)}b_{n^{\prime}}^{(k)},\\
W_{nn^{\prime}}^{(2)}(t)  &  =\sum_{k=1}^{N}\frac{1}{\sqrt{\mu_{k}}}%
\sin\left(  \nu\sqrt{\mu_{k}}t\right)  b_{n}^{(k)}b_{n^{\prime}}^{(k)},\\
W_{nn^{\prime}}^{(3)}(t)  &  =\sum_{k=1}^{N}\sqrt{\mu_{k}}\sin\left(  \nu
\sqrt{\mu_{k}}t\right)  b_{n}^{(k)}b_{n^{\prime}}^{(k)}.
\end{align*}
The operators $q_{n}(t)$ and $p_{n}(t)$ describe the quantum motion of phonons
in the crystal. Under real experimental conditions, the typical order of the
largest frequency $\nu\sqrt{\mu_{N}}$ of the phonon oscillation is
$O(10^{6})~\operatorname*{Hz}$ for $N=2\sim10$, and the typical order of the
ion crystal size is $O(10^{-6})~\operatorname*{m}$. Therefore, the ratio of
the phonon velocity to the light velocity is estimated to be $O(10^{-8})\ll1$.
This estimation ensures that the system can be treated non-relativistically.

In the later discussion, it is convenient to introduce phonon coherent states.
Let us consider a unitary operator given by%
\[
U\left(  \mathbf{\alpha,\beta}\right)  =\exp\left[  i\sum_{n=1}^{N}\left(
\alpha_{n}q_{n}-\beta_{n}p_{n}\right)  \right]  ,
\]
where $\mathbf{\alpha=}\left(  \alpha_{1},\cdots,\alpha_{N}\right)  $ and
$\mathbf{\beta=}\left(  \beta_{1},\cdots,\beta_{N}\right)  $ are real vectors.
By performing $U\left(  \mathbf{\alpha,\beta}\right)  $ on the ground state, a
coherent state is obtained as follows:
\begin{equation}
|\left(  \mathbf{\alpha,\beta}\right)  \rangle=U\left(  \mathbf{\alpha,\beta
}\right)  |g\rangle. \label{25}%
\end{equation}
This state is an eigenstate of the operator $a_{k}$ expressed as
\[
a_{k}|\left(  \mathbf{\alpha,\beta}\right)  \rangle=\left(  \frac{1}%
{\sqrt{2m\nu\sqrt{\mu_{k}}}}A_{k}-i\sqrt{\frac{m\nu\sqrt{\mu_{k}}}{2}}%
B_{k}\right)  |\left(  \mathbf{\alpha,\beta}\right)  \rangle,
\]
where
\begin{align*}
A_{k}  &  =\sum_{n=1}^{N}b_{n}^{(k)}\alpha_{n},\\
B_{k}  &  =\sum_{n=1}^{N}b_{n}^{(k)}\beta_{n}.
\end{align*}
Solving Eqs. (\ref{8}) and (\ref{9}) with $t=0$, $|\left(  \mathbf{\alpha
,\beta}\right)  \rangle$ can be explicitly written as%
\begin{align}
&  |\left(  \mathbf{\alpha,\beta}\right)  \rangle\nonumber\\
&  =\exp\left[  -\frac{1}{4}\sum_{k^{\prime}=1}^{N}\left\vert \frac{1}%
{\sqrt{m\nu\sqrt{\mu_{k^{\prime}}}}}A_{k^{\prime}}-i\sqrt{m\nu\sqrt
{\mu_{k^{\prime}}}}B_{k^{\prime}}\right\vert ^{2}\right] \nonumber\\
&  \times\exp\left[  \sum_{k=1}^{N}\left(  \frac{1}{\sqrt{2m\nu\sqrt{\mu_{k}}%
}}A_{k}-i\sqrt{\frac{m\nu\sqrt{\mu_{k}}}{2}}B_{k}\right)  a_{k}^{\dag}\right]
|g\rangle. \label{26}%
\end{align}
In addition, the inner product between two coherent states is calculated as%
\begin{align}
&  \langle\left(  \mathbf{\alpha,\beta}\right)  |\left(  \mathbf{\alpha
}^{\prime}\mathbf{,\beta}^{\prime}\right)  \rangle\nonumber\\
&  =\exp\left[  \frac{i}{2}\sum_{n=1}^{N}\left(  \alpha_{n}\beta_{n}^{\prime
}-\beta_{n}\alpha_{n}^{\prime}\right)  \right] \nonumber\\
&  \times\exp\left[  -\frac{1}{4}\sum_{k=1}^{N}\left\vert \frac{1}{\sqrt
{m\nu\sqrt{\mu_{k}}}}\left(  A_{k}-A_{k}^{\prime}\right)  -i\sqrt{m\nu
\sqrt{\mu_{k}}}\left(  B_{k}-B_{k}^{\prime}\right)  \right\vert ^{2}\right]  .
\label{27}%
\end{align}
These formulas are used in the next section.

\section{\ \ Quantum Energy Teleportation}

\ 

~

In this section, we analyse a QET protocol by treating the linear ion crystal
discussed in the previous section as a QET channel. The first ion at
$x=x_{1}^{(0)}$ is the gateway of the QET channel where energy is input. The
$N$-th ion at $x=x_{N}^{(0)}$ is the exit of the QET channel where the
teleported energy is output. First, let us perform a local POVM measurement of
the phonon fluctuation in the ground state $|g\rangle$ at the gateway. The
probe qubit used to measure the fluctuation is composed of two internal energy
levels of the gateway ion in the same way as $\cite{HRB}$. The measurement
interaction in our model is given by
\begin{equation}
H_{m}=g(t)\sigma_{y}G_{1},\label{10}%
\end{equation}
where $g(t)$ is a time-dependent real coupling constant, $\sigma_{y}$ is the
$y$ component of the Pauli matrix, and $G_{1}$ is a Hermitian local operator
on the first ion defined by%
\begin{equation}
G_{1}=\phi+\lambda q_{1}.\label{11}%
\end{equation}
Here, $\phi$ and $\lambda$ are time-independent real coupling constants. Let
us assume that the coupling constant $g(t)$ does not vanish only during a very
short period of time via the laser pulse field and that it can be approximated
as
\[
g(t)=\delta(t).
\]
The initial state of the probe qubit is assumed to be the up eigenstate of the
$z$ component of the Pauli matrix, $\sigma_{z}$, given by
\[
|+\rangle=\left[
\begin{array}
[c]{c}%
1\\
0
\end{array}
\right]  .
\]
After the switch off of the measurement interaction in Eq. (\ref{10}), we
projectively measure $\sigma_{z}$ of the probe qubit. This POVM measurement
can be described by two measurement operators \cite{nc} given by%
\begin{equation}
M_{\pm}=\langle\pm|\exp\left[  -i\sigma_{y}G_{1}\right]  |+\rangle,\label{12}%
\end{equation}
where $|-\rangle$ is the down eigenstate of $\sigma_{z}$ given by
\[
|-\rangle=\left[
\begin{array}
[c]{c}%
0\\
1
\end{array}
\right]  .
\]
The operators $M_{\pm}$ are computed explicitly as
\begin{align}
M_{+} &  =\cos G_{1},\label{14}\\
M_{-} &  =\sin G_{1}.\label{15}%
\end{align}
In addition, they satisfy the following relations:
\begin{align}
\sum_{s=\pm}M_{s}^{\dag}M_{s} &  =1,\label{16}\\
\sum_{s=\pm}sM_{s}^{\dag}M_{s} &  =\cos\left(  2G_{1}\right)  .\label{22}%
\end{align}
The average state of the phonon system after the measurement is given by%
\[
\rho_{M}=\sum_{s=\pm}M_{s}|g\rangle\langle g|M_{s}^{\dag}.
\]
Before the measurement, the phonon system is in the ground state $|g\rangle$
and has no energy such that%
\[
\langle g|H|g\rangle=0.
\]
After the measurement, the average energy of the phonon system is given by
\begin{equation}
E_{in}=\operatorname*{Tr}\left[  H\rho_{M}\right]  =\sum_{s=\pm}\langle
g|M_{s}^{\dag}HM_{s}|g\rangle.\label{13}%
\end{equation}
This can be easily evaluated using Eq. (\ref{16}). Taking into account that
$M_{s}$ commutes with the kinetic energy of the ions for $n=2\sim N$ and the
potential energy of all the ions, it can be proven that only the kinetic
energy $p_{1}^{2}/\left(  2m\right)  $ of the gateway ion changes during the
measurement. In order to evaluate this change, the following relation can be
used:$\ $
\begin{equation}
\left[  p_{1},~M_{\pm}\right]  =\pm i\lambda M_{\mp}\label{17}%
\end{equation}
From Eq. (\ref{17}), we can obtain the following relation given by%
\begin{equation}
M_{\pm}^{\dag}p_{1}^{2}M_{\pm}=M_{\pm}^{\dag}M_{\pm}\left(  p_{1}^{2}%
+\lambda^{2}\right)  \mp2i\lambda M_{+}M_{-}p_{1}.\label{19}%
\end{equation}
Using Eqs. (\ref{16}) and (\ref{19}), $E_{in}$ is computed as%
\begin{equation}
E_{in}=\sum_{s=\pm}\langle g|M_{s}^{\dag}\frac{p_{1}^{2}}{2m}M_{s}%
|g\rangle-\langle g|\frac{p_{1}^{2}}{2m}|g\rangle=\frac{\lambda^{2}}%
{2m}.\label{50}%
\end{equation}
This energy of the excited phonons must be infused from outside the system and
is regarded as the energy inputted at the gateway of the QET channel. The
obtained information $s$ is transferred through a classical channel from the
gateway point at $x=x_{1}^{(0)}$ to the exit point at $x=x_{N}^{(0)}$. In
principle, the speed at which the information is transferred is equal to the
speed of light, which is considerably greater than that of phonon
propagation\ in the ion crystal. It is emphasized that the phonons excited at
the QET gateway do not arrive at the exit point when the information arrives
at the exit point. Because the measurement process at the gateway is local,
the state of the exit ion before the arrival of the phonons is locally the
same as the ground state. Interestingly, however, we are able to soon extract
energy from the exit ion by using the transferred $s$. Let us perform a local
unitary operation dependent on $s$ given by%
\begin{equation}
U_{s}=\exp\left(  is\theta p_{N}\right)  ,\label{41}%
\end{equation}
where $\theta$ is a real parameter fixed later. It is possible to suppress the
zero-point oscillation of the phonon fluctuation at the exit ion by performing
the above operation. Neglecting time evolution of the system during the
high-speed transfer of $s$, the average state of the system after the
operation is given by
\begin{equation}
\rho_{F}=\sum_{s=\pm}U_{s}M_{s}|g\rangle\langle g|M_{s}^{\dag}U_{s}^{\dag
}.\label{20}%
\end{equation}
Using Eq. (\ref{13}) and a formula given by%
\[
U_{s}^{\dag}HU_{s}=H-s\theta m\nu^{2}\sum_{n=1}^{N}A_{Nn}q_{n}+\theta^{2}%
\frac{m\nu^{2}}{2}A_{NN},
\]
the average energy after the operation can be evaluated as follows:%
\begin{align}
E_{F} &  =\operatorname*{Tr}\left[  H\rho_{F}\right]  =\sum_{s=\pm}\langle
g|M_{s}^{\dag}U_{s}^{\dag}HU_{s}M_{s}|g\rangle\nonumber\\
&  =E_{in}-\theta m\nu^{2}\sum_{n=1}^{N}A_{Nn}\langle g|q_{n}\left(
\sum_{s=\pm}sM_{s}^{\dag}M_{s}\right)  |g\rangle+\theta^{2}\frac{m\nu^{2}}%
{2}A_{NN}.\label{74}%
\end{align}
Here, we have used the fact that $\left[  M_{s}^{\dag},~q_{n}\right]  =0$.
Substituting Eq. (\ref{22}) into Eq. (\ref{74}) yields%
\begin{equation}
E_{F}=E_{in}-\theta\eta+\theta^{2}\xi,\label{23}%
\end{equation}
where $\eta$ and $\xi$ are real coefficients given by
\begin{align}
\eta &  =m\nu^{2}\sum_{n=1}^{N}A_{Nn}\langle g|q_{n}\cos\left(  2G_{1}\right)
|g\rangle,\label{33}\\
\xi &  =\frac{m\nu^{2}}{2}A_{NN}.\label{34}%
\end{align}
The expression of $\eta$ in Eq. (\ref{33}) can be simplified further as
follows. First, it is pointed out that the following relation holds:
\begin{align}
\cos\left(  2G_{1}\right)  |g\rangle &  =\frac{1}{2}\left[  e^{2iG_{1}%
}+e^{-2iG_{1}}\right]  |g\rangle\nonumber\\
&  =\frac{1}{2}e^{2i\phi}e^{2i\lambda q_{1}}|g\rangle+\frac{1}{2}e^{-2i\phi
}e^{-2i\lambda q_{1}}|g\rangle.\label{102}%
\end{align}
It is observed that the two states $|\pm2\lambda\rangle=e^{\pm2i\lambda q_{1}%
}|g\rangle$ in the above equation are the following phonon coherent states in
Eq. (\ref{25}):%
\[
|\pm2\lambda\rangle=|\left(  \left(  \pm2\lambda,0,\cdots0\right)  ,\left(
0,0,\cdots0\right)  \right)  \rangle.
\]
Therefore, the states $|\pm2\lambda\rangle$ are the eigenstates of $a_{k}$
such that
\begin{equation}
a_{k}|\pm2\lambda\rangle=\pm\frac{2\lambda b_{1}^{(k)}}{\sqrt{2m\nu\sqrt
{\mu_{k}}}}|\pm2\lambda\rangle.\label{31}%
\end{equation}
By introducing a matrix $\Delta_{nn^{\prime}}$ given by
\[
\Delta_{nn^{\prime}}=W_{nn^{\prime}}^{(1)}(0)=\sum_{k=1}^{N}\frac{1}{\sqrt
{\mu_{k}}}b_{n}^{(k)}b_{n^{\prime}}^{(k)}%
\]
and using Eq. (\ref{100}), Eq. (\ref{8}) with $t=0$, and Eq. (\ref{31}), the
following relation is proved.
\begin{equation}
\langle0|q_{n}|\pm2\lambda\rangle=\pm i\frac{\lambda}{m\nu}\langle
0|\pm2\lambda\rangle\Delta_{1n}.\label{80}%
\end{equation}
In addition, Eq. (\ref{27}) yields the following relation:%
\begin{equation}
\langle0|\pm2\lambda\rangle=\exp\left[  -\frac{\lambda^{2}}{m\nu}\Delta
_{11}\right]  .\label{81}%
\end{equation}
From Eq. (\ref{102}), Eq. (\ref{80}), and Eq. (\ref{81}), the following
relation is obtained:%
\[
\langle g|q_{n}\cos\left(  2G_{1}\right)  |g\rangle=-\frac{\lambda\sin\left(
2\phi\right)  }{m\nu}\exp\left[  -\frac{\lambda^{2}}{m\nu}\Delta_{11}\right]
\Delta_{1n}.
\]
Substituting this relation into Eq. (\ref{33}) yields the final expression of
$\eta$ as follows:.
\[
\eta=-\lambda\nu\sin\left(  2\phi\right)  \exp\left[  -\frac{\lambda^{2}}%
{m\nu}\Delta_{11}\right]  \sum_{n=1}^{N}\Delta_{1n}A_{nN}.
\]
From this result, we can show that the coefficient $\eta$ does not vanish as
long as the factor $\sin\left(  2\phi\right)  $ does not vanish. This
observation indicates that the operation in Eq. (\ref{41}) ensures that
$E_{F}$ is smaller than $E_{in}$. In fact, $E_{F}$ in Eq. (\ref{23}) can be
minimized by considering the parameter $\theta$ such that$\,\ $%
\[
\theta=\frac{\eta}{2\xi}.
\]
Then, Eq. (\ref{23}) can be rewritten as%
\begin{equation}
E_{F}=E_{in}-\frac{\eta^{2}}{4\xi}.\label{40}%
\end{equation}
Because $A_{NN}$ in Eq. (\ref{70}) is positive, $\xi$ is also positive. This
implies that during the operation $U_{s}$, a positive amount of energy given
by%
\begin{equation}
E_{out}=\frac{\eta^{2}}{4\xi}=\frac{\lambda^{2}}{2m}\frac{\sin^{2}\left(
2\phi\right)  }{A_{NN}}\exp\left[  -\frac{2\lambda^{2}}{m\nu}\Delta
_{11}\right]  \left\vert \sum_{n=1}^{N}\Delta_{1n}A_{nN}\right\vert
^{2}\label{44}%
\end{equation}
is transferred from the phonon system to the external systems including the
device system executing the operation $U_{s}$. For example, $E_{out}$ for
$N=2$ is analytically evaluated as
\[
E_{out}=\frac{2-\sqrt{3}}{4}\frac{\lambda^{2}}{2m}\sin^{2}\left(
2\phi\right)  \exp\left[  -\left(  1+\frac{1}{\sqrt{3}}\right)  \frac
{\lambda^{2}}{m\nu}\right]  .
\]
This energy extraction is regarded\ as outputting the teleported energy from
the exit of the QET channel. Figure 1 shows the QET protocol. \ Eq. (\ref{44})
can be expressed in terms of the input energy $E_{in}\left(  =\frac
{\lambda^{2}}{2m}\right)  $ as follows:
\begin{equation}
E_{out}=\gamma_{N}E_{in}\exp\left(  -\zeta_{N}\frac{E_{in}}{\nu}\right)
\sin^{2}\left(  2\phi\right)  ,\label{90}%
\end{equation}
where the coefficients $\gamma_{N}$ and $\zeta_{N}\,$\ are defined as
\begin{align}
\gamma_{N} &  =\frac{1}{A_{NN}}\left\vert \sum_{n=1}^{N}\Delta_{1n}%
A_{nN}\right\vert ^{2},\\
\zeta_{N} &  =4\Delta_{11}.
\end{align}
The maximum value of the output energy $E_{out}$ with respect to $\phi$ is
obtained by setting $\phi=\pm\frac{\pi}{4}$. From Eq. (\ref{90}), it is
observed that $E_{out}$ exponentially decays when $\,$the input energy
$E_{in}$ is considerably greater than the typical energy $\nu$ of one phonon
in the ion crystal. Therefore, $E_{in}$ should be of the same order as $\nu$;
this is ensured by choosing $\lambda=O\left(  \sqrt{m\nu}\right)  $ in Eq.
(\ref{11}). Further, it is observed that $E_{out}$ rapidly decays when $N$
becomes large. In Figures 2 and 3, numerical results of $\ln\gamma_{N}$ and
$\zeta_{N}$ are plotted as functions of $N$. Though $\zeta_{N}$ does not have
a drastic $N$-dependence, $\gamma_{N}$ decays approximately exponentially
($\propto e^{-1.1N}$). Hence, the output energy $E_{out}$ behaves as
$e^{-1.1N}O\left(  \nu\right)  $. It is emphasized that without using the
transferred $s$, we cannot extract energy from the exit ion on an average by
performing an arbitrary local quantum operation on the ion. This is because
the state of the exit ion is locally the same as the ground state. Due to the
passivity property of the ground state, any local quantum operation
independent of $s$ on the exit ion gives energy to the crystal, generating an
excited state, or it ensures that the state is unchanged with no energy gain.
The non-negativity of $H$ ensures that $E_{F}\geq0$ in Eq. (\ref{40}). Thus,
$E_{out}$ is not greater than $E_{in}$.

\section{\ \ Summary and Discussion}

\ ~

In this study, we have analysed a QET protocol for the linear ion crystal
model developed by James \cite{James}. At the gateway of the QET channel (at
$x=x_{1}^{(0)}$), the phonon fluctuation in the ground state is POVM measured.
The measurement operators are given by Eqs. (\ref{14}) and (\ref{15}). During
the measurement process, the input energy $E_{in}$ in Eq. (\ref{50}) is
infused from outside the system. The kinetic energy of the gateway ion
increases after the measurement; however, the kinetic energy of the other ions
and the potential energy of all the ions remain unchanged. The obtained
information $s$ by the POVM measurement is transferred through a classical
channel to the exit point at $x=x_{N}^{(0)}$. In principle, the speed at which
the information is transferred is equal to the speed of light, which is
considerably greater than that of the phonon propagation\ in the ion crystal.
Even when the phonons excited at the gateway point do not arrive at the exit
point, we are able to soon extract the energy $E_{out}$ in Eq. (\ref{44}) from
the exit ion by performing $U_{s}$ in Eq. (\ref{41}). Because the intermediate
ions of the crystal are not excited during all the operations of the protocol,
this QET involves the transportation of energy without the generation of heat
in the channel.

Thus far, even though the QET mechanism is an interesting phenomenon, it has
not yet been experimentally verified. Clearly, the experimental verification
of QET requires fast performance of local operations and classical
communication. Therefore, it may be better to perform these operations
collectively using one physical carrier of information. To realize such a
situation, polarization of a laser pulse in a fibre as a probe qubit can be
carried out instead of using the internal energy levels of the ion as the
qubit. The laser pulse is coupled sequentially with the first and $N$-th ions
in the crystal during its propagation in the fibre. This situation is depicted
in Figure 4. To make the discussion more concrete, let us consider the
creation and annihilation bosonic operators $\Psi_{s}^{\dag}(\zeta)$ and
$\Psi_{s}(\zeta)$ for one photon of the laser field with polarization $s=\pm$
in the fibre parametrized by a coordinate $\zeta$. The fibre is connected
between the initial point $\left(  \zeta=\zeta_{i}\right)  $and the final
point $\left(  \zeta=\zeta_{f}\right)  $ via the first ion of the crystal at
$\zeta=\zeta_{1}$ and the $N$-th ion of the crystal at $\zeta=\zeta_{N}$,
where $\zeta_{i}<\zeta_{1}<\zeta_{N}<\zeta_{f}$. The operators $\Psi_{s}%
^{\dag}(\zeta)$ and $\Psi_{s}(\zeta)$ satisfy the following commutation
relations:%
\begin{align*}
\left[  \Psi_{s}(\zeta),~\Psi_{s^{\prime}}^{\dag}(\zeta^{\prime})\right]   &
=\delta_{ss^{\prime}}\delta\left(  \zeta-\zeta^{\prime}\right)  ,\\
\left[  \Psi_{s}(\zeta),~\Psi_{s^{\prime}}(\zeta^{\prime})\right]   &  =0,\\
\left[  \Psi_{s}^{\dag}(\zeta),~\Psi_{s^{\prime}}^{\dag}(\zeta^{\prime
})\right]   &  =0.
\end{align*}
The vacuum state $|0\rangle$ of the laser field is defined by%
\[
\Psi_{s}(\zeta)|0\rangle=0.
\]
Let us assume that the initial state of the laser field is a pulse-wave
coherent state with polarization $s=+$ given by $|f_{i}\rangle\propto
\exp\left(  \int f_{i}(\zeta)\Psi_{+}^{\dag}(\zeta)d\zeta\right)  |0\rangle$,
where $f_{i}(\zeta)$ is a function with a support localized around
$\zeta=\zeta_{i}$. Let us consider a free Hamiltonian of the fibre photon as%
\[
H_{\Psi}=-\frac{ic}{2}\int_{-\infty}^{\infty}\left[  \Psi(\zeta)^{\dagger
}\partial_{\zeta}\Psi(\zeta)-\partial_{\zeta}\Psi(\zeta)^{\dagger}\Psi
(\zeta)\right]  d\zeta,
\]
where $c$ is the light velocity and $\Psi(\zeta)$ is given by
\[
\Psi(\zeta)=\left[
\begin{array}
[c]{c}%
\Psi_{+}(\zeta)\\
\Psi_{-}(\zeta)
\end{array}
\right]  .
\]
The free evolution of the photon field is given by%
\[
e^{itH_{\Psi}}\Psi_{s}^{\dag}(\zeta)e^{-itH_{\Psi}}=\Psi_{s}^{\dag}%
(\zeta-ct).
\]
In order to simulate the protocol discussed in section 3, this laser field
couples with the first and the $N$-th ions by interactions given by%
\begin{align}
H_{M}  &  =\frac{c}{d}G_{1}\int_{\zeta_{1}-d/2}^{\zeta_{1}+d/2}\Psi^{\dagger
}(\zeta)\sigma_{y}\Psi(\zeta)d\zeta,\label{60}\\
H_{LO}  &  =-\frac{\theta c}{d}p_{N}\int_{\zeta_{N}-d/2}^{\zeta_{N}+d/2}%
\Psi^{\dagger}(\zeta)\sigma_{z}\Psi(\zeta)d\zeta, \label{61}%
\end{align}
where $d$ is the length of the interaction region. The total Hamiltonian of
the composite system is expressed as%
\[
H_{tot}=H+H_{M}+H_{LO}+H_{C},
\]
and it is independent of time. Hence, $H_{tot}$ is conserved in time. The
initial state of the composite system is given by $|g\rangle_{q}\otimes
|f_{i}\rangle_{\Psi}$. In this model, the evolution of the laser pulse induces
effective switching of the interactions for the QET operations on the phonon
system. The interactions in Eqs. (\ref{60}) and (\ref{61}) are active only
when the pulses exist at their corresponding interaction regions. Based on the
interaction in Eq. (\ref{60}), the information $s$ about the phonon
fluctuation is imprinted into the polarization of the laser pulse. The
interaction in Eq. (\ref{61}) gives a controlled operation gate of the exit
ion by the value $s$ of $\sigma_{z}$ of the laser pulse. The energy of the
laser pulse changes when the pulse passes through the ion regions. The initial
energy of the pulse is denoted by $E_{1}$. In principle, the photon energy
$E_{1}$ is chosen to be independent of $E_{out}$. In order to perform a
sensitive experiment for the detection of the QET effect, it is better to
consider that the order of $E_{1}$ is the same as that of $E_{out}%
(=O(\gamma_{N}\nu))$. After the interaction with the first ion in Eq.
(\ref{60}), the pulse energy is decreased to $E_{2}$ by the exciting phonons
with energy $E_{in}=E_{1}-E_{2}$ in the crystal. After the interaction with
the $N$-th ion in Eq. (\ref{61}), the pulse energy is increased to $E_{3}$ by
extracting the energy from the ion as the QET effect. The amount of the
extracted energy is $E_{out}=E_{3}-E_{2}$. Experimental observation of this
change in the pulse energy leads to the verification of the QET mechanism.
Detailed analysis of this model will be reported elsewhere.

\bigskip{}

\bigskip{} \textbf{Acknowledgments}

I would like to thank M. Ozawa and A. Furusawa for fruitful discussions. This
research is partially supported by the SCOPE project of the MIC and the
Ministry of Education, Science, Sports and Culture of Japan, No. 21244007.

\bigskip{}

\bigskip{}

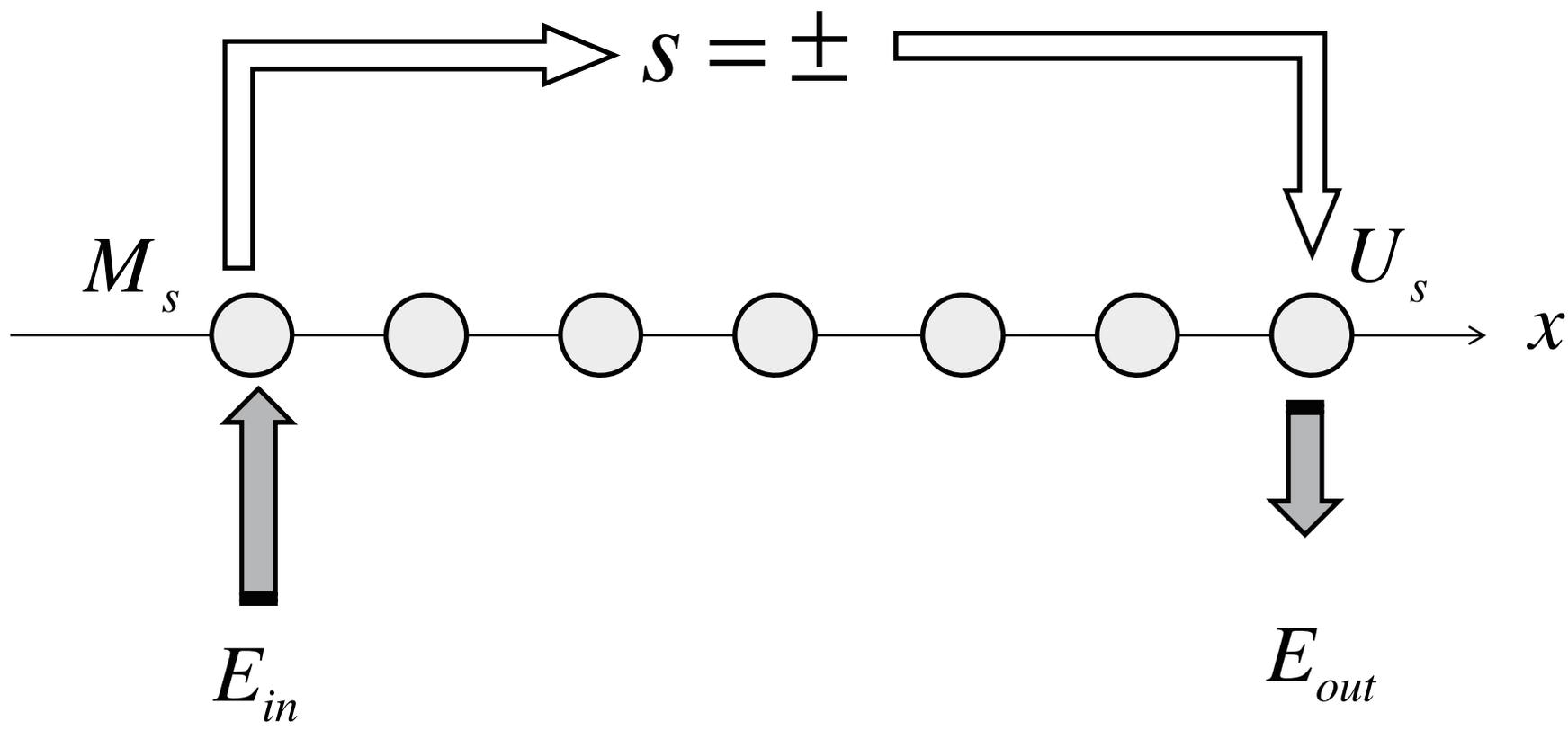
Figure 1: Schematic diagram of QET for the linear ion crystal. The trapped
cold ions are denoted by circles. The POVM measurement defined by Eqs.
(\ref{14}) and (\ref{15}) is performed on the ion at the left edge of the
crystal to obtain the information $s$ about phonon fluctuation in the ground
state $|g\rangle$. During the measurement, energy $E_{in}$ is infused into the
crystal. The measurement result $s$ is transferred to the right edge of the
crystal through a classical channel. The energy $E_{out}$ is extracted by
performing $U_{s}$ in Eq. (\ref{41}) on the ion at the right edge.

\bigskip{}

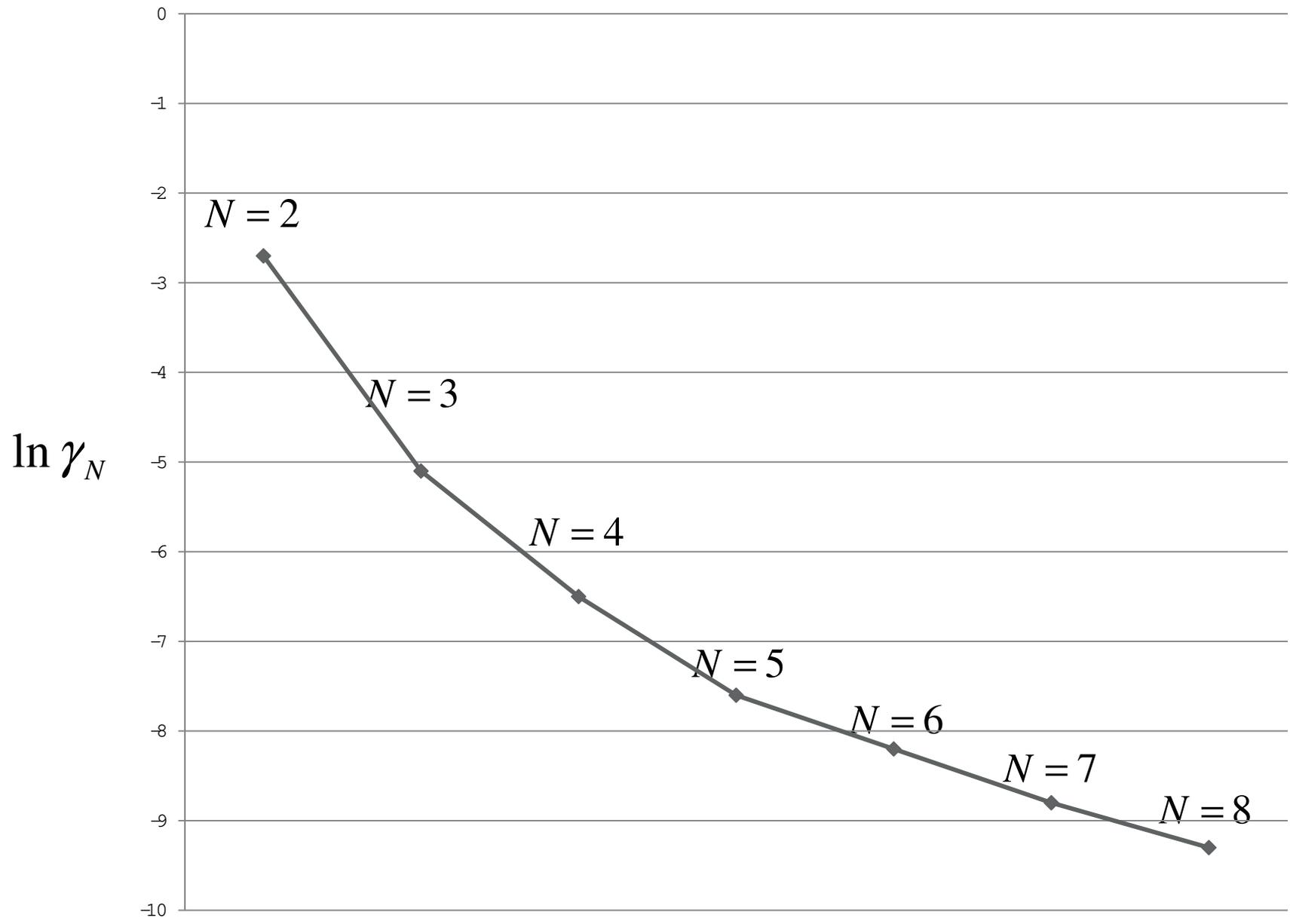
Figure 2: $\ln\gamma_{N}$ is plotted as a function of $N.$

\bigskip{}

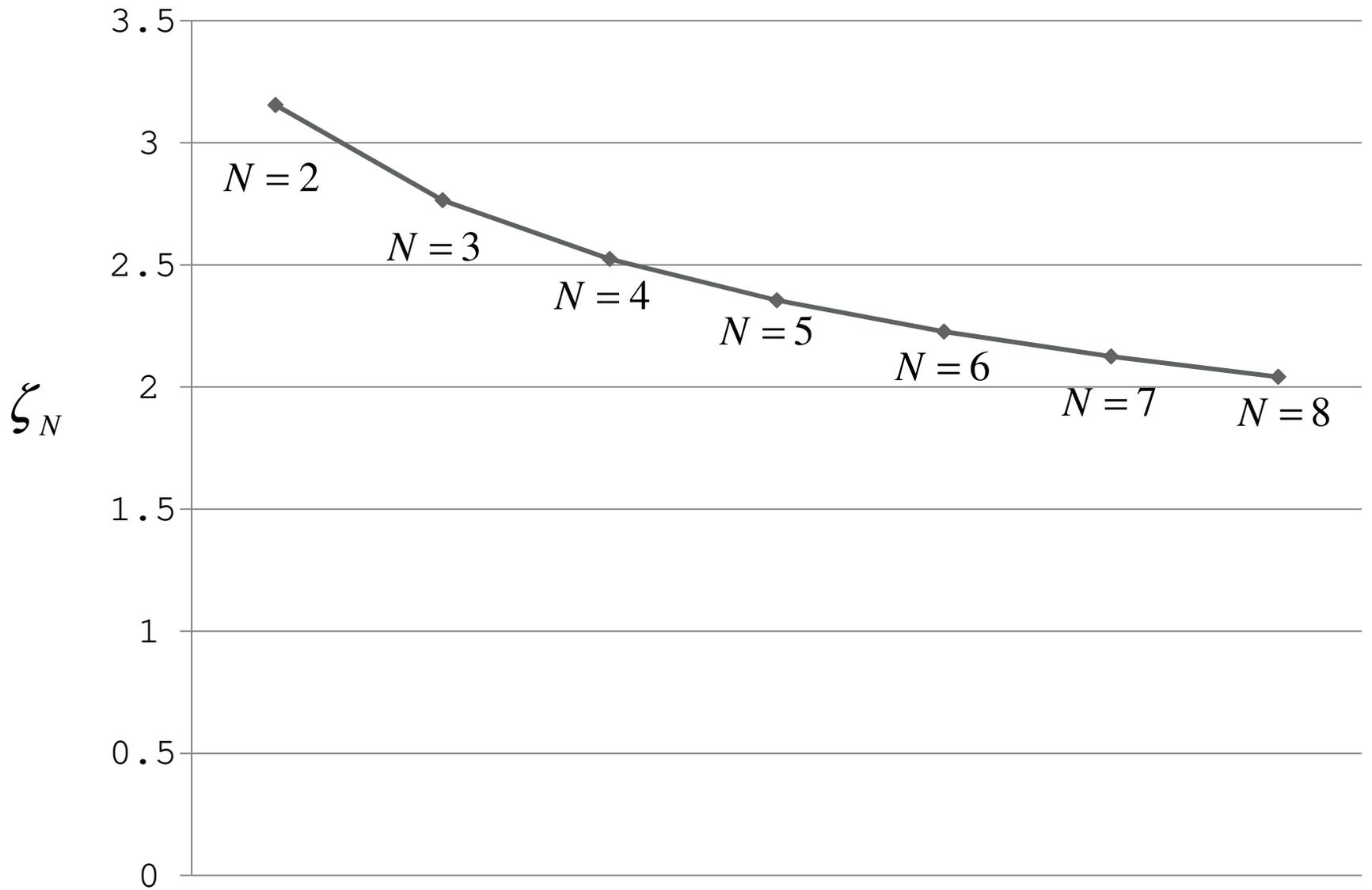
Figure 3: $\zeta_{N}$ is plotted as a function of $N.$

~

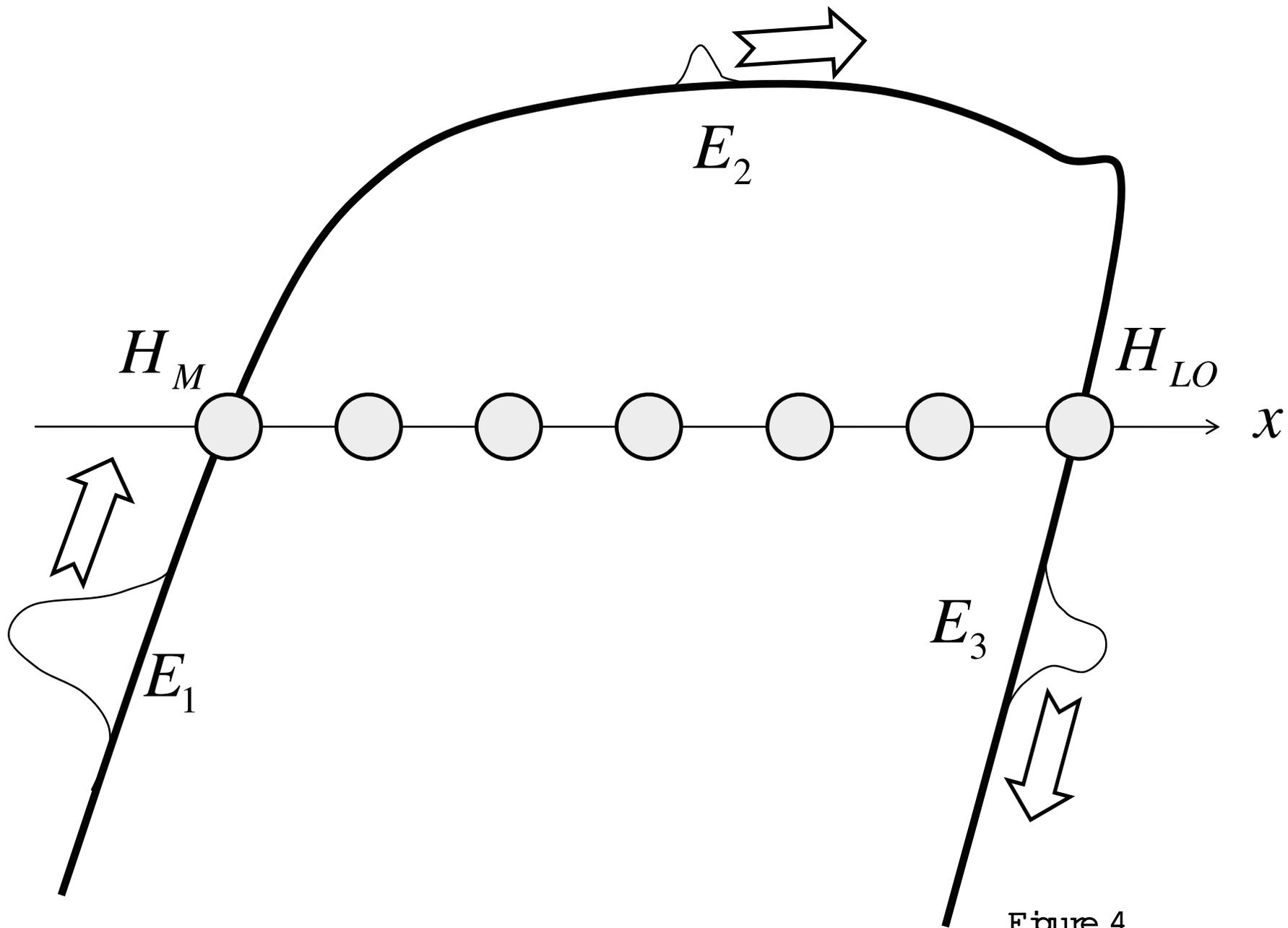
Figure 4: Schematic diagram of the verification experiment of QET. The black
line denotes a photonic fibre for a laser pulse that controls the switching of
the interactions for the QET protocol. The fibre is parametrized by a
coordinate $\zeta$ and connected between the initial point $\left(
\zeta=\zeta_{i}\right)  $ and the final point $\left(  \zeta=\zeta_{f}\right)
$ via the first ion of the crystal at $\zeta=\zeta_{1}$ and the $N$-th ion of
the crystal at $\zeta=\zeta_{N}$. The initial energy of the laser pulse is
denoted by $E_{1}$. After the interaction with the first ion in Eq.
(\ref{60}), the pulse energy is decreased to $E_{2}$ by the exciting phonons
with energy $E_{in}=E_{1}-E_{2}$ in the crystal. After the interaction with
the $N$-th ion in Eq. (\ref{61}), the pulse energy is increased to $E_{3}$ by
extracting the energy from the ion as the QET effect. The amount of energy
extracted is given by $E_{out}=E_{3}-E_{2}$.

\end{document}